\documentclass[twocolumn,prl,aps,superscriptaddress]{revtex4-1}
\usepackage[latin9]{inputenc}
\usepackage{mathtools}
\usepackage{ulem}
\usepackage{amsbsy}
\usepackage{amstext}
\usepackage[unicode=true,pdfusetitle,
 bookmarks=false,
 breaklinks=false,pdfborder={0 0 1},backref=false,colorlinks=false]
 {hyperref}
\usepackage[dvipsnames]{xcolor}
 \hypersetup{colorlinks=true,citecolor=NavyBlue,filecolor=black,linkcolor=black,urlcolor=black}

\makeatletter

\@ifundefined{textcolor}{}
{%
 \definecolor{BLACK}{gray}{0}
 \definecolor{WHITE}{gray}{1}
 \definecolor{RED}{rgb}{1,0,0}
 \definecolor{GREEN}{rgb}{0,1,0}
 \definecolor{BLUE}{rgb}{0,0,1}
 \definecolor{CYAN}{cmyk}{1,0,0,0}
 \definecolor{MAGENTA}{cmyk}{0,1,0,0}
 \definecolor{YELLOW}{cmyk}{0,0,1,0}
}

\usepackage{amsmath}
\usepackage{graphicx}
\usepackage{amssymb}
\usepackage{txfonts,color}
\usepackage{dsfont}
\makeatother

\newcommand{\beq}{\begin{equation}}
\newcommand{\eeq}{\end{equation}}
\newcommand{\beqa}{\begin{eqnarray}}
\newcommand{\eeqa}{\end{eqnarray}}

\begin{document}
\title{Speeding up quantum perceptron via shortcuts to adiabaticity}

\author{Yue Ban}
\email{ybanxc@gmail.com}
\affiliation{Department of Physical Chemistry, University of the Basque Country UPV/EHU, Apartado 644, 48080 Bilbao, Spain}
\affiliation{School of Materials Science and Engineering, Shanghai University, 200444 Shanghai, China}

\author{Xi Chen}
\affiliation{Department of Physical Chemistry, University of the Basque Country UPV/EHU, Apartado 644, 48080 Bilbao, Spain}
\affiliation{International Center of Quantum Artificial Intelligence for Science and Technology (QuArtist) \\and Department of Physics, Shanghai University, 200444 Shanghai, China}

\author{E. Torrontegui}
\affiliation{Instituto de F\'{i}sica Fundamental IFF-CSIC, Calle Serrano 113, 28006 Madrid, Spain}

\author{E. Solano}
\affiliation{Department of Physical Chemistry, University of the Basque Country UPV/EHU, Apartado 644, 48080 Bilbao, Spain}
\affiliation{International Center of Quantum Artificial Intelligence for Science and Technology (QuArtist) \\and Department of Physics, Shanghai University, 200444 Shanghai, China}
\affiliation{IKERBASQUE, Basque Foundation for Science, Plaza Euskadi 5, 48009 Bilbao, Spain}
\affiliation{IQM, Munich, Germany}

\author{J. Casanova}
\affiliation{Department of Physical Chemistry, University of the Basque Country UPV/EHU, Apartado 644, 48080 Bilbao, Spain}
\affiliation{IKERBASQUE, Basque Foundation for Science, Plaza Euskadi 5, 48009 Bilbao, Spain}

\keywords{quantum perceptron, shortcuts to adiabaticity, qubit operation}

\begin{abstract}
The quantum perceptron is a fundamental building block for quantum machine learning. This is a multidisciplinary field that incorporates abilities of quantum computing, such as state superposition and entanglement, to classical machine learning schemes. Motivated by the techniques of shortcuts to adiabaticity, we propose a speed-up quantum perceptron where a control field on the perceptron is inversely engineered leading to a rapid nonlinear response with a sigmoid activation function. This results in faster overall perceptron performance compared to quasi-adiabatic protocols, as well as in enhanced robustness against imperfections in the controls.
\end{abstract}

\maketitle

\section*{Introduction}
In the era of information expansion, the merge of quantum information and artificial intelligence will have a transformative impact in science, technology, and our societies~\cite{quantum-supremacy, quantum-ML1, quantum-ML2}. 
In particular, classical networks of artificial neurons (or nodes) represent a successful framework for machine learning strategies, with the {\it perceptron} being the simplest example of a node~\cite{neural-network}. The perceptron is based on the McCulloch-Pitts neuron~\cite{McCulloch-Pitts}, and it was originally proposed by Rosenblatt in 1957 to create the first trained networks~\cite{perceptron}. Nowadays, extensions of these original ideas such as multilayer perceptrons in networks with interlayer connectivity are exploited to deal with demanding computational tasks.

The emergence of quantum computing and machine learning has boosted the development of both fields \cite{Q-computing1,Q-computing2,Q-computing3,Q-computing4,Q-computing5,Q-computing6,Q-measurement}, giving rise to the field of quantum machine learning.
In this context, quantum neural networks (QNNs) have attracted growing interest~\cite{QNN1, QNN2} since the seminal idea proposed by Kak~\cite{Kak}. In particular, the entering of classical machine learning techniques into the quantum domain has the potential to accelerate the performance of different applications such as classification and pattern recognition~\cite{quantum-ML1,NN-quantum,quantum-perceptron-binary,quantum-supervised-NV,quantum-NN-CV,quantum-NN-hopfield,quantum-NN-swapgate,QCNN-image}.  In addition, nowadays the excellent degree of quantum control over the registers in modern quantum platforms \cite{LeibfriedEtAl, Devoret13, Bloch05, Obrien09} allows the performance of quantum operations with high fidelity, which further feeds the idea of having reliable QNNs. However, the linear and unitary framework of quantum mechanics raises a serious dilemma, since neural networks present nonlinear and dissipative behaviours which are hard to reproduce at the quantum level. 
To address this challenge, many efforts have been attempted by exploiting quantum measurements \cite{Kak,Zak}, the quadratic kinetic term to generate nonlinear behaviours~\cite{Bebrman}, dissipative \cite{Kak} or repeat-until-success \cite{quantum-circuit-Cao} quantum gates, and reversible circuits \cite{Wan}. Among them, gate-based QNNs \cite{QNN-gate} with training optimization procedures  \cite{QNN-gate-training} are feasible to implement by a set of unitary operations. Furthermore, gate-based QNNs can behave as variational quantum circuits that encode highly nonlinear transformations while remaining unitary~\cite{quantum-NN-CV}. Also, a quantum algorithm implementing the quantum version of a binary-valued perceptron was introduced in Ref.~\cite{quantum-perceptron-binary}, showing an exponential advantage in resources storage. Remarkably, a universal {\it quantum perceptron} has been proposed as an efficient approximator in Ref.~\cite{quantum-perceptron}, where the quantum perceptron is encoded in an Ising model with a sigmoid activation function. In particular, the sigmoid nonlinear response is parametrized by the potential exerted by other neurons, and driven by adiabatic techniques.

In this article, motivated by the nonadiabatic control provided by shortcuts to adiabaticity (STA) techniques ~\cite{STA1,STA2}, we design fast sigmoidal responses with the aid of the invariant-based inverse engineering (IE) \cite{IE1,IE2,IE3}. The IE method is based on dynamical modes of Lewis-Riesenfeld invariant instead of one instantaneous eigenstates of the original reference Hamiltonian \cite{Berry,CD-driving}. As IE directly imposes boundary conditions in the wave function evolution, the nonlinear activation function of the quantum perceptron encoded in the probability of the excited state can be achieved in a fast and robust way. In particular, an external control field on the perceptron is designed such that it leads to a fast nonlinear activation function with a wide tolerance window to the variation of the input potential induced by neurons in the previous layer. We demonstrate that our method produces solutions that outperform those based on adiabatic techniques, which significantly facilitates the implementation of quantum perceptrons in modern platforms such as nitrogen vacancy (NV) centers in diamond. Note that, the latter are settings where external control fields can be introduced with extraordinary precision~\cite{Zopes17}.

\section*{Results}
\subsection*{Quantum Perceptron}
\begin{figure}[]
	\begin{center}
		\scalebox{0.35}[0.35]{\includegraphics{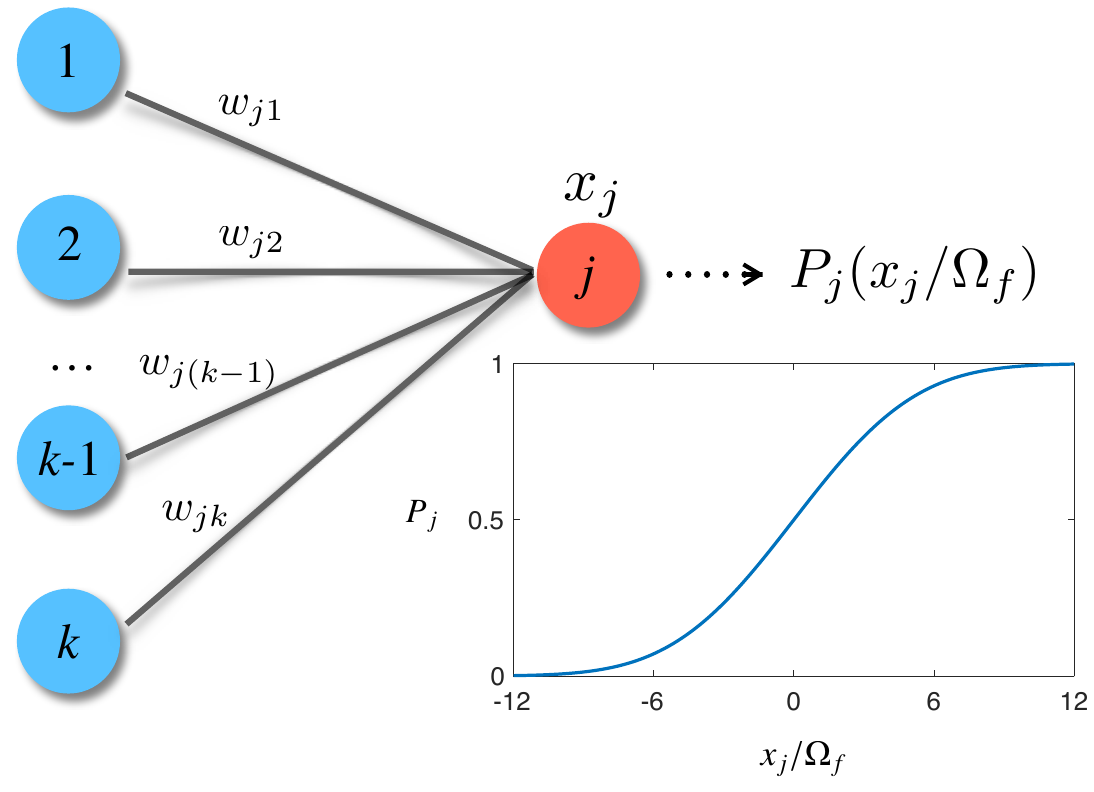}}
		\caption{\label{schematic}  Schematic configuration of a quantum perceptron. When it is integrated in a feed-forward neural network, the potential depends on neurons in earlier layers, e.g., $\hat{x}_j = \sum_{i=1}^k w_{ji} \hat{\sigma}^z_{i} - b_j$, where the activation function of the quantum perceptron is the probability of the excited state $P_j (x_j / \Omega_f)$  at the final time $t=t_f$ in the form of sigmoid-shape, shown in the inset.
		}
	\end{center}
\end{figure}

The capacity of feed-forward neural networks to classify complex data relies in the ``universal approximation theorem" proved by Cybenko \cite{NN2-transfer-line}, claiming that any continuous function can be written as a linear combination of sigmoid functions. A QNN is also demonstrated as a universal approximator of continuous functions \cite{quantum-perceptron}.
In a classical network, a perceptron (or neuron) generates the signal $s_j = f(x_j)$ as a sigmoidal response to the weighted sum of the signals (or outputs) from the neurons in the previous layer. More specifically, $x_j = \sum_{i=1}^k w_{ji} s_i - b_j$ with the neuron interconnectivities $w_{ji}$, the bias $b_j$, and $s_i$ being the output of the $i$th neuron in the previous layer. 
In analogy with classical neurons, a quantum perceptron can be constructed as a qubit that encodes the nonlinear response to an input potential in the excitation probability, see Fig. \ref{schematic}. One possibility for the latter is the following gate~\cite{quantum-perceptron}:
\begin{eqnarray}
 \label{gate}
 \hat{U}_j(\hat{x}_j; f) |0_j\rangle = \sqrt{1-f(\hat{x}_j)} |0_j\rangle + \sqrt{f(\hat{x}_j)} |1_j\rangle,
\end{eqnarray}
where, in close similarity with the classical case, we have $\hat{x}_j = \sum_{i=1}^k w_{ji} \hat{\sigma}^z_{i} - b_j$, where $\hat{\sigma}^z_{i}$ is the $z$ Pauli matrix of the $i$th neuron (qubit),  $w_{ji}$ is interaction between the perceptron $j$ and the $i$th neuron in the previous layer, $b_j$ is the bias of the perceptron.
The transformation in Eq.~(\ref{gate}) can be engineered by evolving adiabatically the qubit with the Ising Hamiltonian ($\hbar = 1$)
\begin{eqnarray}
\label{H}
\hat H(t) &=&  \frac{1}{2}\left[\hat{x}_j  \hat{\sigma}^z_j + \Omega(t)\hat\sigma_j^x\right]\nonumber
\\
& = & \frac{1}{2} \left[ \sum_{i=1}^k (w_{ji} \hat{\sigma}^z_i\hat\sigma^z_j) - b_j \hat{\sigma}_j^z + \Omega(t) \hat\sigma_j^x \right],
\end{eqnarray}
where the $j$th qubit (encoding the quantum perceptron) is controlled by an external field $\Omega(t)$, leading to a tunable energy gap in the dressed-state qubit basis $|\pm\rangle$, with $\hat \sigma^x_j|\pm\rangle = \pm |\pm\rangle$. 
When this perceptron is integrated in a feed-forward neural network, the potential depends on the neurons in earlier layers, as the perceptron interacts with other neurons in the previous layer (labeled by $i = 1, ..., k$) via the $x_j$ potential, see Fig. \ref{schematic}. Therefore, the network is encoded in a Hilbert space via the external potential exerted by other neurons.
The Ising Hamiltonian in Eq.~(\ref{H}) has the reduced eigenstate,
\begin{eqnarray}
\label{eigenstate}
|\Phi(x_j/ \Omega(t)) \rangle= \sqrt{1- f(x_j / \Omega(t))} |0\rangle + \sqrt{f(x_j / \Omega(t))} |1\rangle,~~~~
\end{eqnarray}
where $x_j$ now represents the lowest eigenvalue of the operator $\hat{x}_j$, while $f(x)$ corresponds to a sigmoid excitation probability 
\begin{eqnarray}
\label{sigmoid}
f(x) = \frac{1}{2} \left(1+\frac{x}{\sqrt{1+x^2}}\right).
\end{eqnarray}

In order to generate the state on the right side of Eq.~(\ref{gate}), we propose the following strategy:
First, a Hadamard gate is applied to drive the state from $|0\rangle$ to $|+ \rangle = (|0\rangle +|1\rangle) / \sqrt{2}$. Secondly, by appropriately tuning $\Omega(t)$ according to inverse engineering (IE) techniques (to be explained later), the state $|\Psi(0)\rangle = |+\rangle$ evolves to  $|\Psi(t_f)\rangle =|\Phi(x_j/\Omega_f) \rangle$ (up to some phase factor that can be eventually canceled by a phase gate), along with one eigenstate of the Lewis-Riesenfeld invariant of $\hat{H}$, with $|\Phi(x_j/\Omega_f)\rangle$ being the instantaneous eigenstate of $\hat H(t=t_f;\Omega_f)$, and  $\Omega_f \equiv \Omega(t_f)$. 
It is noteworthy to mention that, unlike the fast quasi-adiabatic passage (FAQUAD) approach~\cite{quantum-perceptron}, our method based on IE does not need to achieve the initial condition $\Omega(0) \gg |x_j|$, as it is not required that the initial state meets one eigenstate of $\hat H(0)$. The latter results in a smooth control field $\Omega(t)$ which is easy to be used in experiments.

Another possibility to achieve $|\Psi(t_f)\rangle$ from $|\Psi(0)\rangle$ is by an adiabatic driving  in a Landau-Zener scheme. However, as it is discussed in Ref.~\cite{quantum-perceptron}, this spends long time and may be unfeasible depending on the coherence time of the physical setup that implements the Hamiltonian in Eq.~(\ref{H}). 

\subsection*{Accelerating Quantum Perceptron By IE} 
We adopt the IE method to achieve the 
$|\Psi(0)\rangle\rightarrow|\Phi(x_j/\Omega_f)\rangle$ state transfer with shorter
time than FAQUAD \cite{FAQUAD}. The control field $\Omega(t)$ is then engineered to guarantee that
at the final evolution time $t=t_f$ the qubit excitation probability $P_j(x_j/\Omega_f)$ corresponds
to a sigmoid-like response, i.e. to a mono-valuate $f$ function satisfying 
$\lim\limits_{x\rightarrow -\infty}f(x)\rightarrow 0$ and $\lim\limits_{x\rightarrow \infty}f(x)\rightarrow 1$. 
Since the universality of neural networks does not rely on the specific shape of the sigmoid function \cite{NN1-transfer-line, NN2-transfer-line}, e.g. Eq. (\ref{sigmoid}), 
we quantify the performance of the control field $\Omega(t)$ in the interval $[-x^{\textrm{max}}, x^{\textrm{max}}]$ with the distance $C=2-F_0-F_1$. Here $F_0= |\langle 0 | \Psi(t_f; x_j / \Omega_f=-x^{\textrm{max}}) \rangle|^2$ 
and  $F_1= |\langle 1| \Psi(t_f; x_j / \Omega_f= x^{\textrm{max}})|^2$ characterize how the engineered states 
overlap with $|0\rangle$ and  $|1\rangle$, at $x_j / \Omega_f=-x^{\textrm{max}}$ and 
$x_j / \Omega_f=x^{\textrm{max}}$ respectively. Note that, for a sigmoid-like function, $C\rightarrow 0$,  for $x^{\textrm{max}} \rightarrow\infty$. Meanwhile, in all the numerical results, the activation function is found to be well-behaved, i.e., the function is monotonic and with a sigmoid-like behaviour, $\lim\limits_{x\rightarrow -\infty}f(x)\rightarrow 0$ and $\lim\limits_{x\rightarrow \infty}f(x)\rightarrow 1$.  As we will see later,  our IE technique also provides with robustness with respect to timing errors.

\begin{figure}[t]
\begin{center}
\scalebox{0.32}[0.32]{\includegraphics{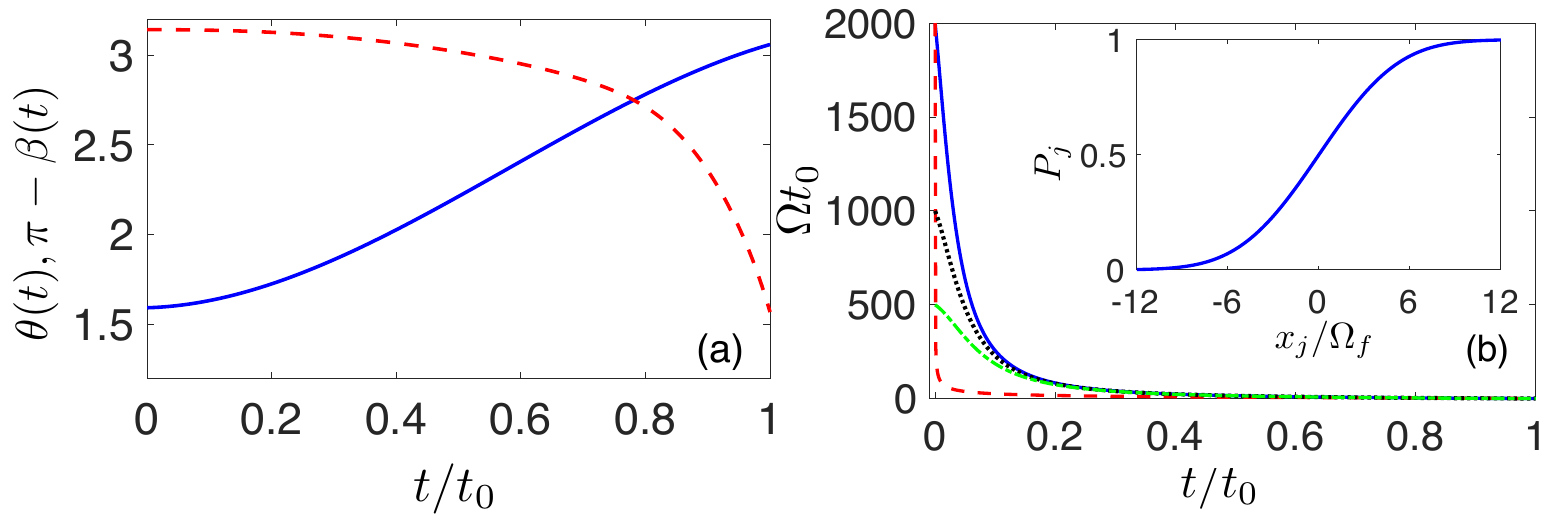}}
\caption{\label{angles-Omega-P} (a) The functions of $\theta$ (solid-blue) and $\pi-\beta$ (dashed-red), where $\theta$ is interpolated by a polynomial ansatz $\theta = \sum_{i=0}^3 a_i t^i$, and $\beta$ is solved from Eq. (\ref{beta}) for  $t_f =1$ and $\kappa=2000$, with $\epsilon=2 \times 10^{-5}$.	
(b) The control fields $\Omega(t)$ designed from IE (solid-blue) with the help of $\theta$, $\beta$ and from FAQUAD (dashed-red), where $\kappa=2000$. We also show the control fields derived from $\kappa =1000$ (dotted-black) and $\kappa=500$ (dot-dashed-green). The inset in (b) displays the corresponding activation functions for different $\kappa$, which coincide with each other. In both plots, $y / \Omega_f =12$. }
\end{center}
\end{figure}

Now we show the procedure to find the control $\Omega(t)$. To this end, we start with the parameterisation of the dynamical state
\begin{eqnarray}
\label{wavefunction}
|\Psi(t)\rangle = \cos(\theta/2) e^{i\beta/2} |0\rangle + \sin(\theta/2) e^{-i\beta/2} |1\rangle,
\end{eqnarray}
with the two unknown polar and azimuthal angles, $\theta \equiv \theta(t)$ and $\beta \equiv \beta(t)$, on the Bloch sphere.
Having the state in Eq.~(\ref{wavefunction}) at hand, the corresponding orthogonal state $|\Psi_{\perp}(t)\rangle$ gets completely determined and the Lewis-Riesenfeld invariant can be thus constructed with constant eigenvalues~\cite{IE1,IE2}.
Substituting one of the states ($|\Psi(t)\rangle $ or  $|\Psi_{\perp}(t)\rangle $) into the time-dependent Schr\"{o}dinger equation driven by the Hamiltonian in Eq.~(\ref{H}), we obtain the following coupled differential equations (for more details see Methods.)
\begin{eqnarray}
\label{Omega}
\Omega(t) &=& \dot{\theta}/\sin \beta,
\\
\label{beta}
x_j  &=&  \dot{\theta} \cot\theta \cot\beta - \dot{\beta}.
\end{eqnarray}
Setting the wavefunction  $|\Psi(0)\rangle=|+\rangle$ and $|\Psi(t_f)\rangle=|\Phi(x_j/\Omega_f) \rangle$ at the initial and final times leads to the boundary conditions
\begin{eqnarray}
\nonumber
\theta(0) &=& 2 \sin^{-1}\left[\sqrt{f({x}_j/ \kappa)}\right], 
\\
\theta(t_f) &=& 2 \sin^{-1}\left[\sqrt{f({x}_j/ \Omega_f)}\right],
\label{theta-conditions-1}
\end{eqnarray}
with the introduced $\kappa$ parameter being infinitely large which results in $|\Phi(x_j/\kappa)\rangle = |+\rangle$. Also, it is important to remark that $\kappa$ does not need to equal the value of our control $\Omega(t)$ at $t=0$,  as $|\Phi(x_j/\kappa)\rangle $ is not necessarily the eigenstate of $\hat H[t=0; \Omega(0)]$. In addition, from Eq. (\ref{Omega}) one can find the following conditions for the first derivatives of $\theta$ at the boundaries
\begin{eqnarray}
\label{theta-conditions-2}
\dot\theta(0) = \Omega(0) \sin \beta(0), \quad \dot\theta (t_f) = \Omega_f \sin \beta(t_f).
\end{eqnarray}
We can interpolate $\theta$ by choosing a simple polynomial function  $\theta = \sum_{i=0}^N a_i t^i $ and a trigonometric fuction $\theta = a_0 + a_1 t +  \sum_{i=2}^N a_i \sin[(i-1)\pi t/t_f]$ with less coefficients required for matching the same boundary conditions \cite{ansatz}.  The appropriate adoptions on the coefficients can make the solution approach the one gained from optimal control theory \cite{davidpra}. 
We present the comparison of the performance of activation function by using IE with these two ansatzes and exponential functions  inspired by regularized optimal solutions in Supplementary Information. We stress that, unlike the method in Ref.~\cite{IE2,IE-annealing}, in our case $\theta$ and $\beta$ are correlated. 
We impose $\beta(t_f) =\pi/2$ and $\beta(0)=\pi-\epsilon$ (note that we will allow a certain deviation by introducing the $\epsilon$ parameter, see later).  Once we construct  $\theta$, the function $\beta$ can be obtained by solving Eq. (\ref{beta}) with the boundary condition $\beta(t_f) =\pi/2$. 
After the functions $\theta$ and $\beta$ are obtained, the control field $\Omega(t)$ is deduced using Eq. (\ref{Omega}).  

The solution to $\beta$ from Eq.~(\ref{beta}) depends on $x_j$ leading to a set of $\Omega\equiv\Omega(t, x_j)$. However, in order to make the control independent of the input potential, we set $\Omega(t)=\Omega(t, x_j=y)$ where the value of $y$ is chosen such that it minimizes the $C$ distance for different $x_j$ in a certain interval (see next section).

\subsection*{IE Performance} 
As the state evolves from $|\Psi(0)\rangle = |+\rangle$, the $\kappa$ parameter should be a large number compared to the input potential $x_j$.
We numerically study situations where $\kappa= 2000$ and explored the range $|x_j|/\Omega_f \in [-x^{\textrm{max}}, x^{\textrm{max}}]$, with $ x^{\textrm{max}}=12$. Note that,  we consider the situation where $x^{\textrm{max}}=12$, although our results are not limited to the specific number. We use dimensionless units, by setting the unit of time $t_0$ such that the control field $\Omega(t)$ is given in terms of $1/t_0$. In addition, we consider an unbiased perceptron with $b_j=0$. 

\begin{figure}[tb]
\begin{center}
\scalebox{0.32}[0.32]{\includegraphics{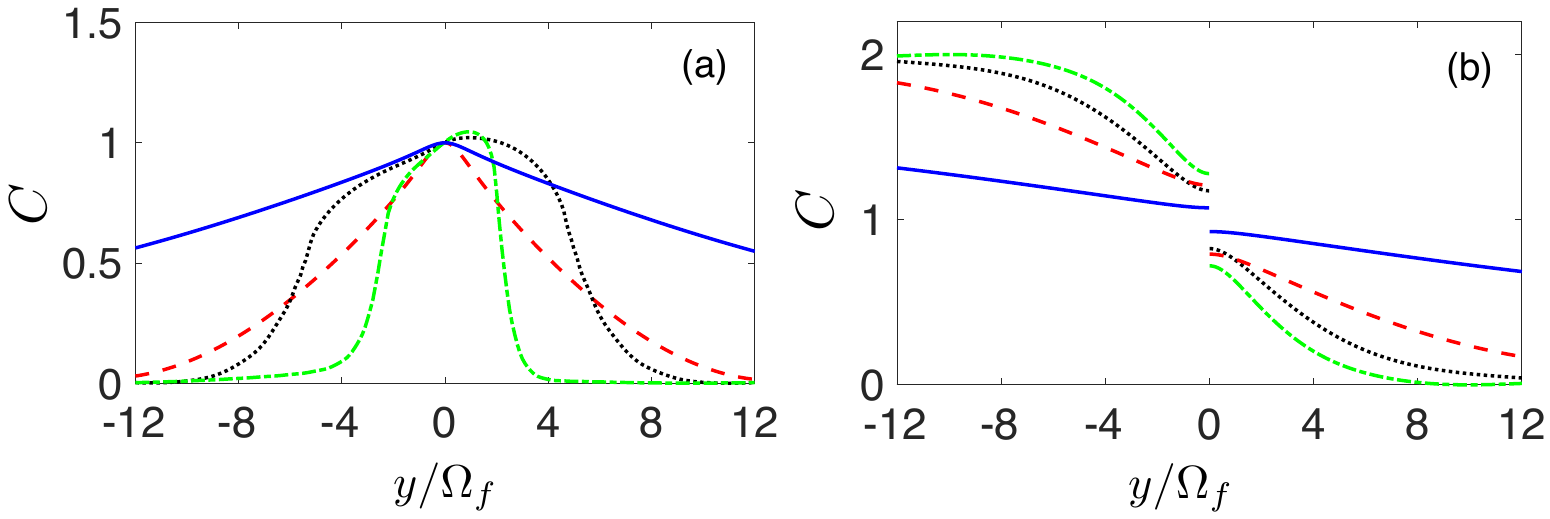}}
\caption{\label{C-y-tf} The dependence of $C$ value on $y$ with the application of IE $\theta =  \sum_{i=0}^3 a_i t^i $ (a) and FAQUAD (b)  for different operation times $t_f = 0.1$ (solid-blue), $t_f=0.2$ (dashed-red), $t_f=0.5$ (dotted-black), and $t_f=1$ (dot-dashed-green). 
}
\end{center}
\end{figure}

Not limited to a fixed large number of $\kappa$, our method shows the flexibility and the feasibility of the control field.
For a case in which we impose $\Omega_f=1$ and solve Eq.~(\ref{beta}) with a fixed value for $x_j/\Omega_f=y/\Omega_f=12$, we find $\theta(0) = 1.576 \simeq \pi/2$. Figure \ref{angles-Omega-P}(a) indicates the obtained solutions for $\theta$ and $\beta$ for this case in which we have also selected the operation time $t_f=1$. We find that the boundary condition for $\beta(0)$ is also satisfied with a tiny error of $\epsilon=2\times 10^{-5}$. In this specific case, we find that the designed control $\Omega(t)$ at $t=0$ is $\Omega(0) = 1999.6 \approx \kappa$ when $\kappa=2000$, the initial state corresponds to the eigenstate state of the Hamiltonian. Also, we observed that $\beta(0)$ tends to $\pi$ when $t_f$ gets larger.
In Fig.~\ref{angles-Omega-P}(b), the control field $\Omega (t)$ obtained with our method is illustrated. This $\Omega (t)$  leads to an excitation probability such that it arrives at $P_j(x^{\textrm{max}}) = 0.998$. Using the same control field $\Omega(t)$, we find that the probability of the state $|1\rangle$ for other input neural potentials $x_j / \Omega_f \in[-x^{\textrm{max}}, x^{\textrm{max}}]$ is in the form of a sigmoid-like response ranging from $0$ to $1$ during the interval, as shown in the inset of Fig. \ref{angles-Omega-P}(b). This proves the successful construction of a sigmoid-shape transfer function, which is a crucial factor for a quantum perceptron. The fields calculated from $\kappa=1000$, $\kappa =500$ lead to the same sigmoid activation function which,  as shown in the inset of Fig. \ref{angles-Omega-P} (b),  cannot be distinguished to the one derived from $\kappa = 2000$.

Our IE method provides a wider range of $y/\Omega_f$ than FAQUAD to construct sigmoid transfer functions. In Figure~\ref{C-y-tf} (a) the value of the distance $C$ obtained with the IE method, as a function of $y/\Omega_f$ for various operation times $t_f$, is shown.
It can be observed that a low value for $C$ appears with large values for $|y|$ and $t_f$. We have checked (also for $t_f = 1$) the appearance of nonlinear perceptron responses that connect $0$ and $1$ with a sigmoid shape. In particular, these lead to $C< 10^{-2}$ in the range $y/\Omega_f  \in [5,12]$ with control fields $\Omega(t)$ for $t_f=1$ similar to the one in Fig. \ref{angles-Omega-P} (b).
In contrast, $C$ goes to almost $2$ at $y / \Omega_f=-x^{\textrm{max}}$ by FAQUAD techniques \cite{FAQUAD}, in which only for long $t_f$ and in the regime $y / \Omega_f \rightarrow x^{\textrm{max}}$  the transfer function can be produced, see Fig.~\ref{C-y-tf} (b). 

\begin{figure}[]
\begin{center}
\scalebox{0.35}[0.35]{\includegraphics{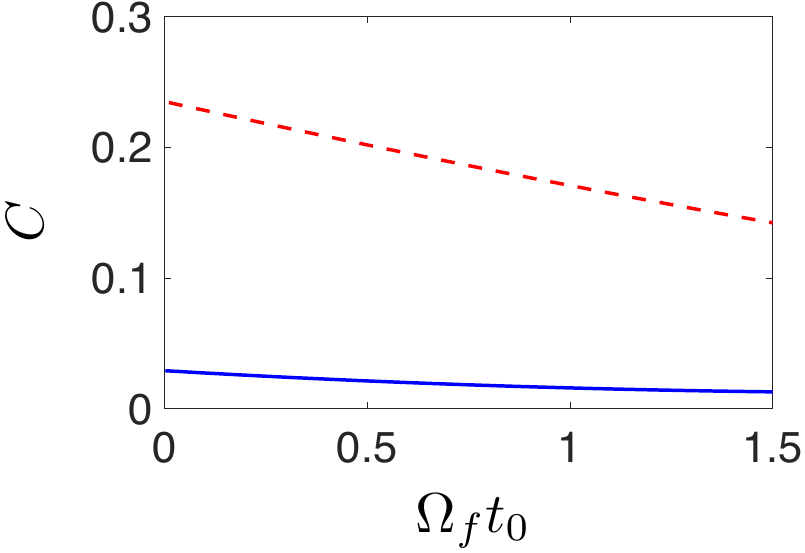}}
\caption{\label{C-Omf} Dependence of $C$ value on $\Omega_f$ is shown for IE $\theta =  \sum_{i=0}^3 a_i t^i $  (solid-blue) and FAQUAD (red-dashed) protocols, when $t_f=0.2$, $y/\Omega_f=12$.
}
\end{center}
\end{figure}

The target state $|\Psi(t_f)\rangle = |\Phi(x_j/\Omega_f)\rangle$ depends on the value of the driving field at the final time, see Eq.~(\ref{eigenstate}). In general we observe that, with our IE method, a larger value of the control field at $t=t_f$ (i.e. $\Omega_f$) offers higher fidelity. As an example of the latter, in Fig. \ref{C-Omf} we show the value of $C$ as a function of $\Omega_f$ for $t_f=0.2$ with the application of IE (solid-blue) and FAQUAD (dashed-red). In this figure one can observe the improved performance of our IE method. Actually, every point of the lower value $C$ by IE implies the successful discovery of sigmoid-shape transfer function and driving field $\Omega(t)$.

\subsection*{Quasi-optimal-time Solution} 
\begin{figure}[tb]
\begin{center}
\scalebox{0.32}[0.32]{\includegraphics{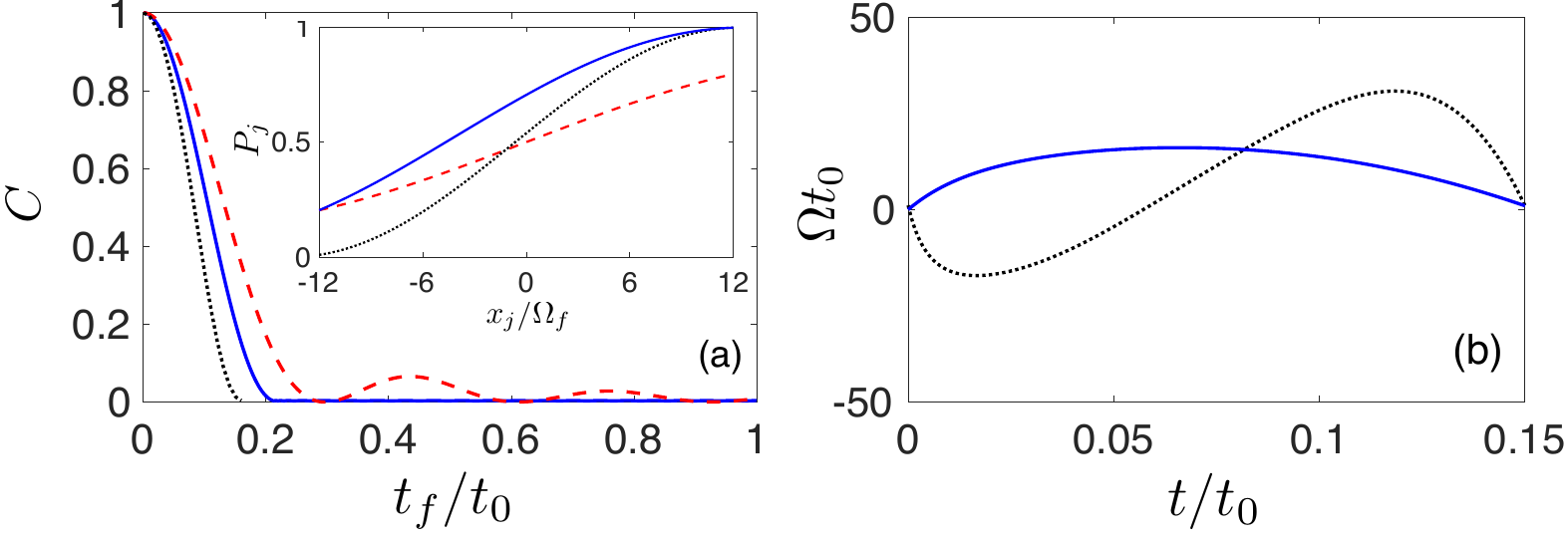}}
\caption{\label{C-tf-P-field} ~(a) Dependence of $C$ as a function of the final time $t_f$, using IE in the cases of $\theta= \sum_{i=0}^3 a_i t^i $ (solid-blue), $\theta= \sum_{i=0}^5 a_i t^i $ (dotted-black) and FAQUAD (dashed-red). The inset of (a) shows the corresponding transfer functions for $t_f =0.15$, where the dotted-black curve represents the quasi-optimal-time solution with $a_2 = -50$, $a_3 = -3980$.
(b) For $t_f=0.15$, the driving field $\Omega(t)$ designed from IE in the cases of using $\theta= \sum_{i=0}^3 a_i t^i $ (solid-blue), using $\theta= \sum_{i=0}^5 a_i t^i $ with the optimal parameters $a_2 = -50$, $a_3 = -3980$ (dotted-black), and $y/\Omega_f =12$.
}
\end{center}
\end{figure}

As the activation function $P(x_j / \Omega_f)$ connects  $0$  and $1$ at $-x^{\textrm{max}}$ and $x^{\textrm{max}}$, we set $C<0.01$ as the criteria of successful construction of a quantum perceptron. In Fig. \ref{C-tf-P-field}(a), we illustrate the dependence of $C$ value on $t_f$ by using the polynomial ansatz  $\theta = \sum_{i=0}^N a_i t^i $ with $N=3$ and $N=5$ of IE as well as FAQUAD~\cite{quantum-perceptron}.
When $N=3$, the smallest $t_f$, such that $C < 0.01$ is satisfied, is $0.2$, while employing techniques based on FAQUAD, this is at $t_f=0.3$.
The further reduction of the smallest $t_f$, such that $C < 0.01$ is satisfied, can be improved since IE method allows to approach the quasi-optimal-time solution by introducing more degrees of freedom in the ansatz of $\theta$ \cite{davidpra}, leading to faster quantum perceptrons. With $N = 5$ (i.e. a solution with two additional parameters, namely $a_4$ and $a_5$), see Fig. \ref{C-tf-P-field}(a) (dotted-black curve) we get a speed up of $2$ with respect to FAQUAD method, leading to the minimal operation time $t_f^{\textrm{min}} = 0.15$.
The values of the transfer function at $-x^{\textrm{max}}$ and $x^{\textrm{max}}$ and $C$ value with the application of IE strategies in polynomial, trigonometric  and exponential functions as well as FAQUAD can be seen in Supplementary Information, showing that high-order polynomial ansatz can give a quasi-optimal-time solution.

Moreover, we find that the IE method is robust with respect to timing errors, i.e. variations on the operation time $t_f$. More specifically, once the minimal value of $C$ is reached for solid-blue in Fig.~\ref{C-tf-P-field}(a), $C$ does not show any appreciable oscillation for $t>t_f^{\textrm{min}}$. Conversely, the FAQUAD driving leads to the dashed-red curve in Fig.~\ref{C-tf-P-field}(a) that shows an oscillatory behavior of $C$, 
indicating that only at some specific $t_f$ the sigmoid transfer function can be constructed.

Remarkably, for short times, e.g. $t_f=0.15$, the transfer functions and driving fields are completely different for IE and FAQUAD protocols. In the inset of Fig. \ref{C-tf-P-field} (a) and in Fig. \ref{C-tf-P-field} (b), we give the detailed demonstration of transfer functions and driving fields designed from IE. On the one hand, FAQUAD protocol cannot produce the sigmoid function, by connecting from $0$ to $1$ at the edges, see the inset of Fig. \ref{C-tf-P-field} (a) dashed-red curve. On the other hand, we find that the case of IE with the polynomial ansatz of $N=3$ fails to connect the state $|0\rangle$ presenting $P(-x^{\textrm{max}}) = 0.2$ (solid-blue curve). However, We can overcome this limitation by increasing the order of the polynomial ansatz to $N=5 $. 
Here, we compare the activation functions achieved by different strategies at the same value of $y/\Omega_f=12$. It is worth mentioning that by increasing the value of $x^{\textrm{max}}$ which means more energy is supplied to the system, we can recover a more stretched sigmoid with the FAQUAD protocol or IE with the polynomial ansatz of $N=3$. However, in this work,  we find the external driving $\Omega(t)$ by which the perceptron can have a sigmoidal response in a fixed Hamiltonian configuration with the range $[-x^{\textrm{max}}, x^{\textrm{max}}]$.

In addition, the derived controls $\Omega(t)$ from IE methods are smooth and present values close to zero at $t=0$, see Fig. \ref{C-tf-P-field}(b). Compared to the case of $t_f=1$, shorter operation time leads to larger $\epsilon$ so that $\Omega(0)$ is farther away from $\kappa$.  This is in contrast with the control $\Omega(t)$ derived from FAQUAD techniques that demands an abrupt change from $\Omega(0) = 2000$ to $\Omega(t_f) = 1$, see  Fig. \ref{angles-Omega-P} (b). This demonstrates the appropriateness of our IE derived controls  to be implemented experimentally. In this regard, in the next section we give estimations based on state of the art experimental parameters in NV centers in diamond that demonstrates the suitability of an implementation of our method in such quantum platform.

\section*{Discussions}
We have demonstrated that the enhanced performance of our  method using IE techniques leads to sigmoid activation functions within a minimal operation time of $t_f^{\textrm{min}}=0.15  \ t_0$.  If, for instance, one selects $t_0 = 500 $ ns, the maximum value for the control $\Omega(t)$ amounts to $|\Omega_{\rm max}| \approx 50$ MHz for the kind of solutions presented in Fig.~\ref{C-tf-P-field} (b) (see horizontal axis limits in that figure). This permits the application of our controls in modern quantum platforms such as NV centers in diamond that present coherence times much longer than $0.15  \ t_0=0.15 \times 500 =  75$ ns even at room temperature~\cite{Doherty13, Dobrovitski13}.  In addition, current arbitrary waveform generators allow to change the amplitude of the delivered microwave field (and consequently of the Rabi frequency $\Omega$) in time-scales significantly smaller than $1$ns~\cite{Zopes17, NaydenovPC}. Then, one can easily introduce the controls in Fig.~\ref{C-tf-P-field} (b) to produce nonlinear sigmoid responses in NV centers. 
IE is also helpful to achieve the robust control in a specific physical setup \cite{STA-unwanted-transitions1,STA-unwanted-transitions2,STA-unwanted-transitions3} when one considers the Ising model with unwanted transitions between the target two-level system and other levels.
In this manner one could envision a diamond chip with several NVs, each of them with available nearby nuclear spin qubits, as a quantum hardware to construct QNN using IE methods.

\section*{Methods}
\subsection*{Inverse Engineering And Derivation Of Auxiliary Differential Equations}
The quantum perceptron gate evolves a qubit with the general Hamiltonian (Eq. (\ref{H}))
which has the instantaneous ground state (Eq. (\ref{eigenstate}))
with the basis $|0\rangle= (0,1)^T$ and $|1\rangle = (1,0)^T$ and a sigmoid excitation probability (Eq. (\ref{sigmoid})). Therefore, we need to control the final state exactly as $|\Psi(t_f)\rangle = |\Phi(x_j / \Omega(t_f))\rangle$ in the form of Eq. (\ref{eigenstate}). 
Inverse engineering by parameterizing the Bloch sphere angles $\theta$ and $\beta$ can manipulate the dynamical state evolution in a fast way. After substituting the wave function $|\Psi(t)\rangle$ (Eq. (\ref{wavefunction})) or the orthogonal state $|\Psi_{\perp}(t)\rangle$ into Schr\"{o}dinger equation, we can obtain two equations
\begin{eqnarray}
\label{A1}
-i  \dot{\theta}\sin \frac{\theta}{2} -  \dot{\beta}\cos \frac{\theta}{2} &=& x_j \cos \frac{\theta}{2} + \Omega(t) \sin \frac{\theta}{2} e^{-i \beta},
\\
\label{A2}
i  \dot{\theta} \cos \frac{\theta}{2} +  \dot{\beta}\sin \frac{\theta}{2} &=& -x_j \sin \frac{\theta}{2} +\Omega(t) \cos \frac{\theta}{2} e^{i \beta}.
\end{eqnarray}
Eq. (\ref{A1}) $\times \sin(\theta/2)$ $+$ Eq. (\ref{A2}) $\times \cos(\theta/2)$ and  Eq. (\ref{A1}) $\times \sin(\theta/2)$ $-$ Eq. (\ref{A2}) $\times \cos(\theta/2)$, respectively, result in the analytical expressions of $\Omega(t)$ (Eq. (\ref{Omega})) and $\beta$ (Eq. (\ref{beta})). 
Once setting the operation time $t_f$ and the dynamics of the polar angle $\theta$, we can obtain the function $\beta$ by solving Eq. (\ref{beta}) with the boundary condition $\beta(t_f) = \pi/2$. 
Hence, from Eq. (\ref{Omega}), we derive the applied field $\Omega(t)$.

\subsection*{Fast Quasi-Adiabatic Method}
Another protocol to construct a quantum perceptron by controlling the qubit gate is to use FAQUAD strategy \cite{quantum-perceptron, FAQUAD}, which can achieve the fast and adiabatic-like procedure.
The adiabatic parameter 
\begin{eqnarray}
\label{adiabatic-parameter}
\mu(t) = \hbar \left| \frac{\langle \phi_0(t) | \partial_t \phi_1(t) \rangle}{E_1(t) - E_0(t)} \right|
\end{eqnarray}
is kept as a constant $\mu(t) =c$ during the whole control process, where the instantaneous eigenstates for the Hamiltonian (Eq. \ref{H})
are 
\begin{eqnarray}
|\phi_l\rangle = \cos(\alpha/2) |1\rangle +(-1)^l \sin(\alpha/2) |0\rangle
\end{eqnarray}
with the eigenenergies are~$E_l = - (-1)^l \hbar\sqrt{\Omega^2 + x^2_j} /2$,~$\alpha = \arccos\left[-x_j / \sqrt{\Omega^2+x^2_j}\right]$~and $l \in \{0,1\}$.
In order to construct a universal quantum gate, a single control should not depend on the neuron potential $x_j$. The largest value $|\mu|$ occurs at $|x_j / \Omega_f| \approx 1.272$.  We take this $\mu$ value as an optimal condition that works for all input neuron configurations. As the relation between the field and time is invertible, we can apply the chain rule to Eq. (\ref{adiabatic-parameter}) and obtain
\begin{eqnarray}
\frac{d \Omega}{d t} = - \frac{c}{\hbar} \left| \frac{E_1(\Omega) - E_0(\Omega)}{\langle \phi_0(\Omega) | \partial_\Omega \phi_1 (\Omega) \rangle} \right|,
\end{eqnarray}
where the negative sign represents $\Omega(t)$ monotonously decreases from $\Omega(0)$ to $\Omega(t_f)$. The total duration time is rescaled as $s = t/t_f$ so that $\tilde\Omega(s) := \Omega(s ~ t_f)$ and $d\Omega / dt =t_f^{-1} d\tilde{\Omega} /ds$. As a result, we have 
\begin{eqnarray}
\label{Omegas}
\frac{d \tilde\Omega}{d s} &=& - \frac{\tilde{c}}{\hbar} \left| \frac{E_1 - E_0}{\langle \phi_0 | \partial_{\tilde\Omega} \phi_1\rangle} \right|_{\tilde{\Omega}},
\\
\tilde{c} &= & c t_f = -\hbar \int_{\tilde{\Omega}(0)}^{\tilde{\Omega}(1)} \frac{d \tilde{\Omega}}{\left| \frac{E_1 -E_0}{\langle \phi_0 | \partial_{\tilde\Omega}\phi_1\rangle}\right|_{\tilde{\Omega}}}.
\end{eqnarray}
A selection of $t_f$ corresponds to different scaling of $\tilde{c}$ and $\Omega(t =s t_f) = \tilde{\Omega}(s) $. Consequently, we can derive $\Omega(t)$ from $\tilde{\Omega}(s)$ by solving the differential equation (Eq. (\ref{Omegas})).

\section*{acknowledgements}
We acknowledge financial support from Spanish Government via PGC2018-095113-B-I00 (MCIU/AEI/FEDER, UE), Basque Government via IT986-16, as well as from QMiCS (820505) and OpenSuperQ (820363) of the EU Flagship on Quantum Technologies, and the EU FET Open Grant Quromorphic (828826). J. C. acknowledges the Ram\'{o}n y Cajal program (RYC2018-025197-I) and the EUR2020-112117 Project of the Spanish MICINN, as well as support from the UPV/EHU through the Grant EHUrOPE. X. C. acknowledges NSFC (12075145), SMSTC (2019SHZDZX01-ZX04, 18010500400 and 18ZR1415500), the Program for Eastern Scholar and the Ram\'{o}n y Cajal program of the Spanish MICINN (RYC-2017-22482). E. T. acknowledges support from Project PGC2018-094792-B-I00 (MCIU/AEI/FEDER,UE), CSIC Research Platform PTI-001, and CAM/FEDER Project No. S2018/TCS-4342 (QUITEMAD-CM). 

\section*{Author contributions statement}
 Y. Ban developed the theoretical formalism, performed the analytic calculations and performed the numerical simulations. X. Chen and E. Torrontequi verified the analytical method. J. Casanova supervised the project. All the authors contributed to the final version of the manuscript. 
 
\section*{Additional information}
\textbf{Competing interests}: The authors declare no competing financial interests.

\pagebreak
\widetext
\begin{center}
\textbf{ \large Supplemental Material: \\ Speeding up Quantum Perceptron via Shortcuts to Adiabaticity }
\end{center}

\setcounter{equation}{0} \setcounter{figure}{0} \setcounter{table}{0}
\makeatletter 
\global\long\def\theequation{S\arabic{equation}}
 \global\long\def\thefigure{S\arabic{figure}}
 \global\long\def\bibnumfmt#1{[S#1]}
 \global\long\def\citenumfont#1{S#1}
\renewcommand{\thesection}{\Alph{section}}
\numberwithin{equation}{section}

\section*{Quasi-optimal-time Solution by Inverse Engineering}
\begin{figure}[h]
\begin{center}
\scalebox{0.3}[0.3]{\includegraphics{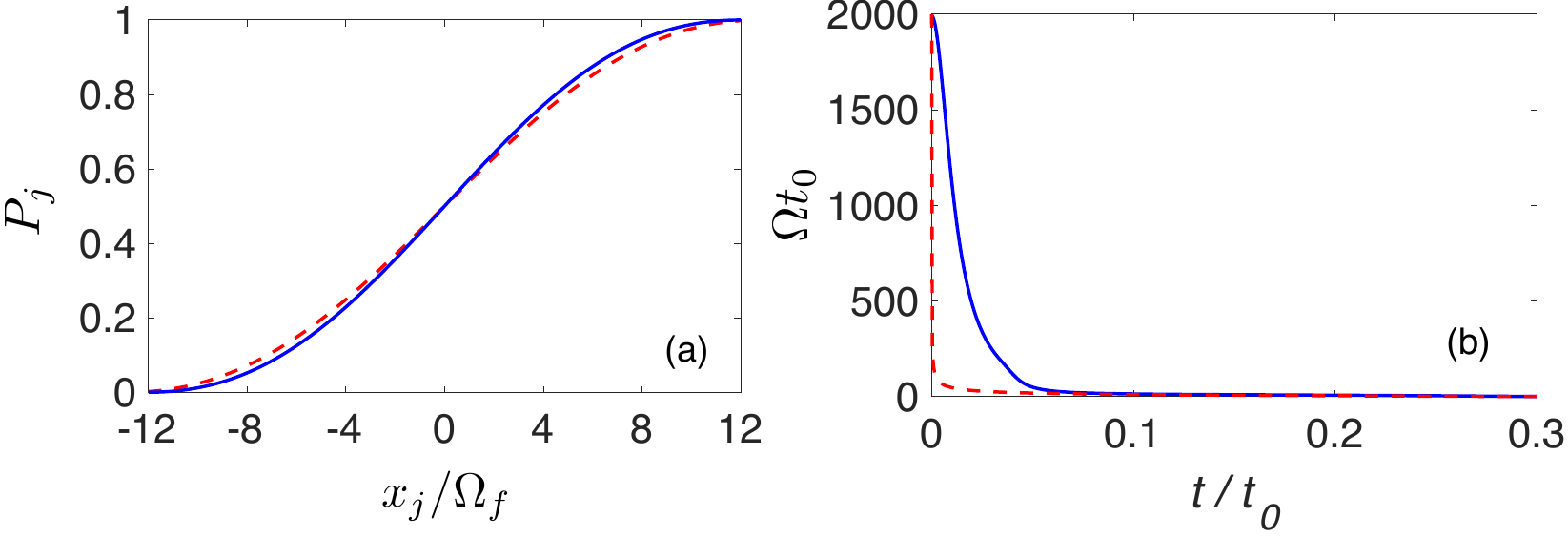}}
\caption{\label{P-field} With $t_f=0.3$, we present the transfer function (a) and the external field $\Omega(t)$ (b) obtained from IE with $\theta= \sum_{i=0}^3 a_i t^i $ (solid-blue) and FAQUAD (dashed-red). In both cases, $\Omega(t)$ is designed when $y/ \Omega_f =12$.
}
\end{center}
\end{figure}

In the main text, we have introduced the inverse engineering (IE) to find the control field and obtain the sigmoid transfer function. Here, we provide the detailed comparison of transfer functions and driving fields between IE and FAQUAD methods for the operation time $t_f = 0.3$, see Fig. \ref{P-field}.
The transfer functions for both IE in the case of $\theta= \sum_{i=0}^3 a_i t^i$ and  FAQUAD protocols can reach $1$ and $0$ at $x_j / \Omega_f=x^{\textrm{max}}$ and $x_j/\Omega_f=-x^{\textrm{max}}$ ($x^{\textrm{max}} = 12$) with high fidelity, respectively.  However, the driving field $\Omega(t)$
for IE decreases more smoothly from the 
maximum value $\Omega(0) = 1999.5$ for $\kappa=2000$, which makes the experimental implementation more feasible.

We clarify the manner of  doing quasi-optimal-time control as follows. The coefficients of the polar angle $\theta= \sum_{i=0}^s a_i t^i$ with $s=3$ can be solved from the boundary conditions of $\theta(0)$, $\theta(t_f)$, $\dot\theta(0)$, $\dot\theta(t_f)$ for a fixed value $t_f$. The polar angle can also be set into a higher order polynomial ansatz ($s> 3$), where the unknown free coefficients can be scanned to seek for a lowest $C$ value.

\begin{figure}[b]
\begin{center}
\scalebox{0.4}[0.4]{\includegraphics{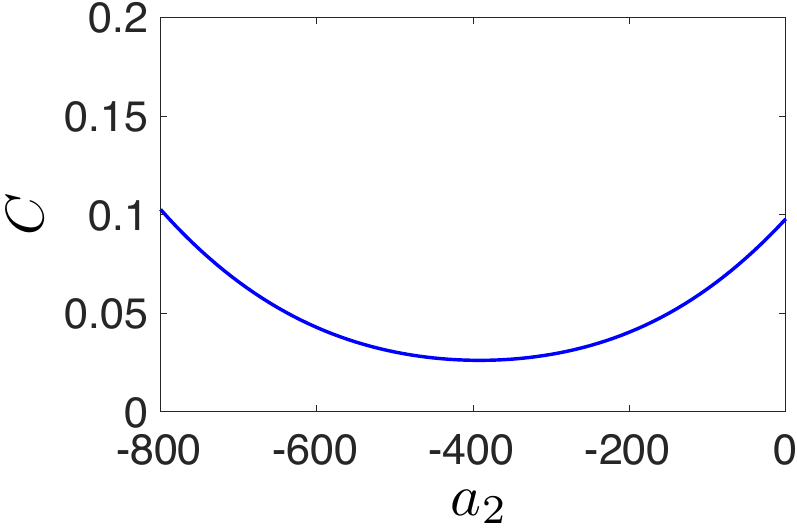}}
\caption{\label{IE-1parameter} With $t_f = 0.15$, the dependence of $C$ value on the free parameter $a_2$, where $\theta = \sum_{i=0}^4 a_i t^i$, and $y / \Omega_f =12$.}
\end{center}
\end{figure}

\begin{figure}[t]
\begin{center}
\scalebox{0.4}[0.4]{\includegraphics{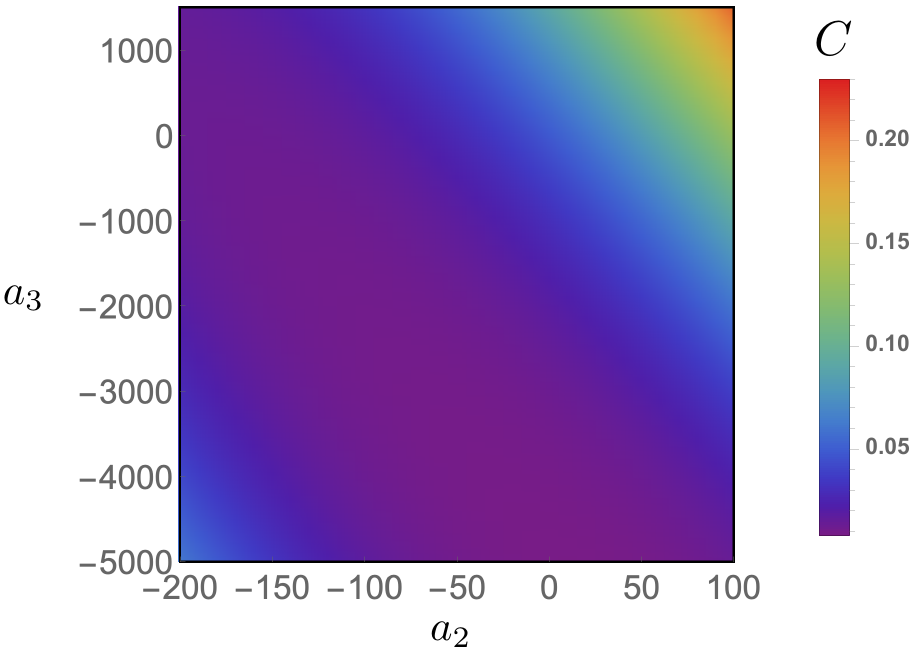}}
\caption{\label{IE-2parameter} With $t_f = 0.15$, the dependence of the density contour plot of $C$ on the free parameters $a_2$ and $a_3$, where  $\theta = \sum_{i=0}^5 a_i t^i$ and $y/ \Omega_f =12$.}
\end{center}
\end{figure}

For $t_f=0.15$, we first set $s=4$, and obtain $a_0 = \theta(0)$, $a_1 = \dot\theta(0)$, $a_3$ and $a_4$ and the functions of $a_2$ by fixing the boundary conditions Eq. (\ref{theta-conditions-1}) and Eq. (\ref{theta-conditions-2}) in the main text. 
As shown in Fig. \ref{IE-1parameter}, the minimum $C= 0.026$ can be found at $a_2=-391$.
By using the same boundary conditions, we set $s=5$, a higher order polynomial ansatz, where $a_0 = \theta(0)$, $a_1 = \dot\theta(0)$, $a_4$ and $a_5$ are the functions of $a_2$ and $a_3$. The relation of $C$ value versus $a_2$ and $a_3$ are demonstrated in Fig. \ref{IE-2parameter}, where the range of $C<0.01$ manifests itself as a stripe area. We find numerically
$C$ value reaches its minimum at $0.0087$ when $a_2 = -50$ and $a_3 = -3980$.
Using the same strategy to search for a minimal $C$ value for a fixed value $t_f$, we demonstrate $C$ value in the function of $t_f$, as shown in Fig. \ref{C-tf-P-field} (a) of the main text, where the minimal operation time $t_f$ reaches at $t_f^{\textrm{min}}=0.15$ for $C<0.01$. Numerical calculations prove that further setting higher order of polynomial ansatz ($s>5$) does not improve to shorten  $t_f^{\textrm{min}}$.

The detailed comparison between STA and optimal control theory is presented in Ref. \cite{davidpra-S}, proving that IE method allows to approach the performance gained from optimal control theory by introducing more freedom in polynomial or trigonometric ansatz of $\theta$. Here, we present the comparison of the performance of activation function by using IE with polynomial function $\theta = \sum_{i=0}^N a_i t^i $, trigonometric function $\theta = a_0 + a_1 t +  \sum_{i=2}^N a_i \sin[(i-1)\pi t/t_f]$ and exponential functions  $\theta = a_0 e^{t} + a_1 e^{-t} + a_2 e^{m t} +a_3 e^{-m t}$ with $m=25$ as well as FAQUAD, shown in Table \ref{Comparison}, showing that higher polynomial ansatz gives a quasi-optimal-time solution.

\begin{table}[]
	\centering
	\begin{tabular}{lcccc}
		\hline
		& $N$ order & $P(-x^{\textrm{max}})$ & $P(x^{\textrm{max}})$ & $C$ \\
		\hline
		Polynomial & 3 & 0.204 & 0.998 & 0.206 \\
		& 4 & 0.024 & 0.998 & 0.026 \\
		& 5 & 0.0065 & 0.998 & 0.008\\
		\hline
		Trigonometric & 2 & 0.219 & 0.998 & 0.221\\
		& 3 & 0.0534 & 0.998 & 0.0554\\
		& 4 & 0.0429 & 0.998 & 0.0389\\
		\hline
		Exponential &  & 0.086 & 0.998 & 0.0878\\
		\hline
		FAQUAD & & 0.204 & 0.796 & 0.41\\
		\hline
	\end{tabular}
\caption{\label{Comparison}
Comparison of  the performance of different ansatz: polynomial, trigonometric and exponential functions in form introduced in the main text, with $t_f=0.15$, $x^{\textrm{max}}=12$, $y/\Omega_f =12$.}
\end{table}

\end{document}